\documentclass[aps,twocolumn,a4paper]{revtex4}%
\usepackage{amssymb}
\usepackage{amsmath}
\usepackage{amsfonts}
\usepackage{graphicx}
\usepackage{graphics}

\def\Cas{\mathrm{Cas}}

\def\B{\mathrm{B}}

\def\bk{\mathbf{k}}

\def\dd{\mathrm{d}}

\def\TE{\mathrm{TE}}
\def\TM{\mathrm{TM}}
\def\K{K}
\def\P{\mathrm{P}}    

\def\max{\mathrm{max}}
\def\min{\mathrm{min}}
\def\perf{\mathrm{perf}}
\def\Gold{\mathrm{Gold}}
\def\PFA{\mathrm{PFA}}
\def\Tr{\mathrm{Tr}}
\def\calS{\mathcal{S}}
\def\calR{\mathcal{R}}
\def\calK{\mathcal{K}}
\def\calD{\mathcal{D}}
\def\calF{\mathcal{F}}

\begin{document}

\title{Casimir Effect : Optomechanics in Quantum Vacuum}
\author{Astrid Lambrecht and Serge Reynaud\\
Laboratoire Kastler Brossel, CNRS, ENS, UPMC, Campus Jussieu,
F-75252 Paris, France }

\begin{abstract}
The Casimir effect results from the optomechanical coupling between
field fluctuations and mirrors in quantum vacuum. This contribution
to the 20th International Conference on Laser Spectroscopy (ICOLS
2011) discusses the current status in the comparison between theory
and experiments.
\end{abstract}

\maketitle

\section{Introduction}

The Casimir effect \cite{Casimir48} is a jewel with many facets.
First, it is a macroscopic effect of the irreducible field
fluctuations which fill quantum vacuum. As a crucial prediction of
quantum theory, it has thus been the focus of a number of
works (see reviews in~\cite{Milonni94,LamoreauxAJP99,%
BordagPR01,MiltonJPA05,BarreraNJP06,CasimirPhysics,LambrechtCasimir11}).

Then, it has fascinating interfaces with some of the most important
open questions in fundamental physics. It is connected with the
puzzles of gravitational physics through the problem of vacuum
energy \cite{GenetDark02,JaekelMass08} as well as with the principle
of relativity of motion through the dynamical Casimir-like effects
\cite{JaekelRPP97}. Effects beyond the Proximity Force Approximation
also make apparent the rich interplay of vacuum energy with geometry
\cite{BalianPoincare03,BalianSigrav04,ReynaudJPA08} (more
discussions below).

Casimir physics also plays an important role in the tests of gravity
at sub-millimeter ranges \cite{Fischbach98,AdelbergerARNPS03}.
Strong constraints have been obtained in short range Cavendish-like
experiments \cite{KapnerPRL07}. Should an hypothetical new force
have a Yukawa-like form, its strength could not be larger than that
of gravity if the range is larger than 56$\mu$m. For scales of the
order of the micrometer, gravity tests are performed
by comparing the results of Casimir force measurements with theory%
~\cite{LambrechtPoincare03,OnofrioNJP06,DeccaEPJC07}. Other
constraints can be obtained with atomic or nuclear force
measurements (for a recent overview of short-range tests,
see~\cite{AntoniadisCRAS11}).

Finally, the Casimir force and closely related Van der Waals force
are dominant at micron or sub-micron distances. This entails that
they have strong connections with various active domains and
interfaces of physics, such as atomic and molecular physics,
condensed matter and surface physics, chemical and biological
physics, micro- and nano-technology~\cite{Parsegian06}. In the
following, we will stress that Casimir physics reveals
optomechanical couplings of macroscopic mirrors with quantum vacuum
fields.

\section{The puzzle of vacuum energy}

The classical idealization of space as being absolutely empty was
already affected by the advent of statistical mechanics, when it was
realized that space is in fact filled with black body radiation. The
first quantum law was designed by Planck precisely to explain the
properties of this black body radiation~\cite{Planck00}. In modern
terms, it gave the mean energy per electromagnetic mode as the
product $\overline{n}\hbar\omega$ of the photon energy $\hbar
\omega$ by the mean number of photons per mode
$\overline{n}=\left(\exp\frac{\hbar\omega}{k_\mathrm{B}T}-1\right)^{-1}$.

Like Einstein, Planck was aware of the unsatisfactory character of
his derivation. Among other physicists, they attempted for years to
give more satisfactory treatments by studying in more detail the
interaction between matter and radiation. These attempts led to the
discovery by Einstein of the quantum absorption-emission laws and of
the Bose statistics (see \cite{Wolf79,MilonniAJP91,Sciama91}). In
1911, Planck \cite{Planck12} wrote a new expression for the mean
energy per mode $\left(\overline{n}+\frac12\right)\hbar\omega$ which
contained a zero-point energy $\frac12\hbar\omega$ besides the black
body energy. In contrast to the latter, the zero-point fluctuations
were still present at zero temperature. The arguments thus used by
Planck cannot be considered as consistent today. The first known
argument still acceptable today was proposed by Einstein and
Stern~\cite{Einstein13} in 1913~: the second Planck law (but not the
first one) reproduces the classical limit
$\left(\overline{n}+\frac12\right)\hbar\omega= k_{{\rm B}}T+O\left(
\frac{1}{T}\right)$ at high temperatures $T\to\infty$. Amazingly,
this argument fixes the magnitude of zero-point fluctuations,
essentially visible at low temperatures, by requiring their
disappearance at high temperatures to be as perfect as possible~!

Some physicists took zero-point fluctuations seriously, long before
the advent of the fully developed quantum theory. Debye insisted on
observable consequences of zero-point atomic motions, in particular
through their effects on the intensities of diffraction peaks
\cite{Debye14}. Mulliken produced experimental evidence of the
effects of zero-point motions by studying isotopic shifts in
vibrational spectra of molecules \cite{Mulliken24}. Nernst was the
first physicist to notice, in 1916, that zero-point fluctuations of
the electromagnetic field constituted a challenge for gravitation
theory~\cite{Nernst16,Browne95}. When the energy density is
calculated by summing up the energies over all field modes, a finite
value is obtained for the first Planck law (this is the solution of
the `ultraviolet catastrophe') but an infinite value is produced
from the second law. When a high frequency cutoff $\omega _{{\rm
max}}$ is introduced, the calculated energy density
$\frac{\left( \hbar \omega _{{\rm max}}\right) ^{4}}%
{8\pi^{2}\left( \hbar c\right) ^{3}}$ is finite but still much
larger than the mean energy observed in the world around us through
gravitational phenomena~\cite{WeinbergRMP89}. The ratio of
calculated to observed energy density has a huge value, up to
$10^{120}$ in the most extreme estimations~\cite{AdlerAJP95}.

This major problem, known since 1916 and still unsolved today, has
led famous physicists to deny the reality of vacuum fluctuations. In
particular, Pauli stated in his textbook on Wave
Mechanics~\cite{Pauli33} : \emph{At this point it should be noted
that it is more consistent here, in contrast to the material
oscillator, not to introduce a zero-point energy of
$\frac12\hbar\omega$ per degree of freedom. For, on the one hand,
the latter would give rise to an infinitely large energy per unit
volume due to the infinite number of degrees of freedom, on the
other hand, it would be principally unobservable since nor can it be
emitted, absorbed or scattered and hence, cannot be contained within
walls and, as is evident from experience, neither does it produce
any gravitational field.} A part of these statements is simply
unescapable~: it is just a matter of evidence that the mean value of
vacuum energy does not contribute to gravitation as an ordinary
energy. But it is certainly no longer possible to uphold today that
vacuum fluctuations have no observable effects. Certainly, vacuum
fluctuations are \emph{scattered} by matter, as shown by their
numerous effects in atomic~\cite{CohenTannoudji92} and
subatomic~\cite{Itzykson85} physics. And the Casimir effect is
nothing but the evidence of vacuum fluctuations making their
existence manifest when being \emph{contained within walls}.

\section{The Casimir force}

Casimir calculated the force between a pair of perfectly smooth,
flat and parallel plates in the limit of zero temperature and
perfect reflection. In this idealized case, the expressions for the
force $F_\Cas$ and energy $E_\Cas$ reveal a universal effect
resulting from the confinement of vacuum fluctuations
\begin{eqnarray}
F_\Cas=-\frac{\dd E_\Cas}{\dd L} \quad,\quad E_\Cas= - \frac{\hbar c
\pi^2 A}{720 L^3}
\end{eqnarray}
with $L$ the distance, $A$ the area, $c$ the speed of light and
$\hbar$ the Planck constant. This universality is explained by the
saturation of the optical response of mirrors reflecting 100\% of
incoming fields. In particular the expressions $F_\Cas$ and $E_\Cas$
do not depend on the atomic structure constants.

This idealization is no longer tenable for the real mirrors used in
the experiments. It is thus necessary to take into account the
optical properties of these
mirrors~\cite{LambrechtEPJ00,SvetovoyPRB08}. The most precise
experiments have been performed with metallic mirrors which are good
reflectors only at frequencies smaller than their plasma frequency
$\omega_\P$. Their optical response is described by a reduced
dielectric function written at imaginary frequencies $\omega=i\xi$
as
\begin{eqnarray}
\varepsilon \left[i\xi\right] = \hat{\varepsilon}\left[i\xi\right] +
\frac{\sigma \left[i\xi\right] }{\xi} \quad,\quad \sigma
\left[i\xi\right] = \frac{\omega_\P^2}{\xi+\gamma}
\end{eqnarray}
The function $\hat{\varepsilon} \left[i\xi\right] $ represents the
contribution of interband transitions and is regular at the limit
$\xi\to0$. Meanwhile $\sigma \left[i\xi\right]$ is the reduced
conductivity ($\sigma$ is measured as a frequency and the SI
conductivity is $\epsilon_0\sigma$) which describes the contribution
of the conduction electrons.

A simplified description corresponds to the lossless limit $\gamma
\to 0$ often called the plasma model. As $\gamma$ is much smaller
than $\omega_\P$ for a metal such as Gold, this simple model
captures the main effect of imperfect reflection. However it cannot
be considered as an accurate description since a much better fit of
tabulated optical data is obtained with a non null value of
$\gamma$. When taking into account the imperfect reflection of the
metallic mirrors, one finds that the Casimir force is reduced with
respect to the ideal Casimir expression at all distances for a null
temperature. This reduction is conveniently represented as a factor
$\eta_F = F/F_\Cas$ where $F$ is the real force and $F_\Cas$ the
ideal expression. For the plasma model, there is only one length
scale, the plasma wavelength $\lambda_P = 2\pi c/\omega_P $ (136nm
for Gold). The ideal Casimir formula is recovered ($\eta_F\to1$) at
large distances $L\gg\lambda_P$, as expected from the fact that
metallic mirrors tend to be perfect reflectors at low frequencies
$\omega \ll\omega_P$. At short distances in contrast, a significant
reduction of the force is obtained ($\eta_F\propto L/\lambda_P$), as
a consequence of the fact that metallic mirrors are poor reflectors
at high frequencies $\omega \gg\omega_P$. In other words, there is a
change in the power law for the variation of the force with
distance. This change can be understood as the result of the Coulomb
interaction of surface plasmons living at the two matter-vacuum
interfaces~\cite{GenetAFLB04,IntravaiaPRL05}.

Experiments are performed at room temperature so that the effect of
thermal fluctuations has to be added to that of vacuum
fields~\cite{SchwingerAP78,GenetPRA00}. Significant thermal
corrections appear at distances $L$ larger than a critical distance
determined by the thermal wavelength $\lambda_T$ (a few micrometers
at room temperature). Bostr\"{o}m and Sernelius were the first to
remark that the small non zero value of $\gamma$ had a significant
effect on the force at non null temperatures~\cite{BostromPRL00}. In
particular, there is a large difference at large distances between
the expectations calculated for $\gamma=0$ and $\gamma\neq0$, their
ratio reaching a factor 2 when $L\gg\lambda_T$. It is also worth
emphasizing that the contribution of thermal fluctuations to the
force is opposite to that of vacuum fluctuations for intermediate
ranges $L\sim\lambda_T$. This situation has led to a blossoming of
contradictory papers (see references
in~\cite{ReynaudQfext04,BrevikNJP06,IngoldPRE09}). As we will see
below, the contradiction is also deeply connected to the comparison
between theory and experiments.

Another important feature of the recent precise experiments is that
they are performed in the geometry of a plane and a sphere. The
estimation of the force in this geometry uses the so-called
\textit{Proximity Force Approximation} (PFA)~\cite{DerjaguinQR68}
which amounts to integrating over the distribution of local
inter-plate distances the pressure calculated in the geometry with
two parallel planes. But Casimir forces are certainly not additive~!
The PFA can only be valid when the radius $R$ of the sphere is much
larger than the separation $L$ between the plane and the sphere.
Even in this case its accuracy remains a question of importance for
the comparison between theory and experiments.

We now discuss the status of comparisons between Casimir experiments
and theory. After years of improvement on both sides, we have to
face discrepancies in these comparisons : there are differences
between experimental results and theoretical predictions drawn from
the expected models, as well as disagreements between some recent
experiments.

On one hand, there have been experiments in  Purdue and Riverside
for approximately ten years, the results of which point to an
unexpected conclusion
(see~\cite{DeccaAP05,DeccaPRD07,KlimchitskayaRMP09}). The Purdue
experiment uses dynamic measurements of the resonance frequency of a
microresonator. The shift of the resonance gives the gradient of the
Casimir force in the plane-sphere geometry, which is also (within
PFA) the Casimir pressure between two planes. The typical radius of
the sphere is $R=150\mu$m and the range of distances
$L=0.16-0.75\mu$m. The results appear to fit predictions obtained
from the lossless plasma model $\gamma=0$ rather than those
corresponding to the expected dissipative Drude model $\gamma\neq0$
(see Fig.1 in~\cite{DeccaPRD07}), in contradiction with the fact
that Gold has a finite static conductivity $\sigma_0 =
\omega_\P^2/\gamma$. Note that these experiments are performed at
distances where the thermal contribution as well as the effect of
$\gamma$ are not so large, so that the estimation of accuracy is a
critical issue in these experiments.

On the other hand, there is now a new experiment in
Yale~\cite{SushkovNatPh11}, where a much larger sphere $R=156$mm is
used, allowing for measurements at larger distances $L=0.7-7\mu$m.
The thermal contribution is large there and the difference between
the predictions at $\gamma=0$ and $\gamma\neq0$ is significant.
Another problem appears which is the
large contribution of the electrostatic patch effect%
~\cite{SpeakePRL03,ChumakPRB04,KimPRA10,deManJVST10,LamoreauxCasimir11}.
After subtraction of this contribution of the patch effect, the
results of the Yale experiment fit the expected Drude model. Of
course, these new results have to be confirmed by further
studies~\cite{MiltonNatPh11}.

The conclusion of this discussion is that the Casimir effect, now
measured in several experiments, is however not tested at the 1\%
level, as has been sometimes claimed. In particular, the patch
effect remains a source of concern for Casimir experiments, as for
other precision measurements
(see examples in~\cite{CampJAP91,TurchettePRA07,DeslauriersPRL06,%
RobertsonCQG06,EpsteinPRA07,PollackPRL08,AdelbergerPPNP09,%
EverittPRL11,ReasenbergCQG11}). The patch distribution has not been
measured in any of the experiments discussed above and progress
could of course come with better characterization and control of
surfaces. The related problem of surface roughness has also to be
studied in more
detail~\cite{vanZwolPRB09,BroerEPL11,vanZwolCasimir11}. We note also
that the aspect ratio $L/R$ lies in the range
$[10^{-3},5\times10^{-3}]$ for the Purdue experiment,
$[5\times10^{-5},5\times10^{-6}]$ for the Yale experiment, so that
the corrections to PFA could have quite different effects in the two
cases.

\section{The Casimir effect in the scattering approach}

The best tool available for addressing these questions is the
scattering approach. This approach has been used for years for
evaluating the Casimir force between non perfectly reflecting
mirrors. It is today the best solution for calculating the force in
arbitrary geometries~\cite{LambrechtNJP06}.

The basic idea is that mirrors are described by their scattering
amplitudes. It can be simply illustrated with the model of scalar
fields propagating along the two directions on a line (1-dimensional
space; see references in \cite{JaekelRPP97}). Each mirror is
described by a scattering matrix containing reflection and
transmission amplitudes. Two mirrors form a Fabry-Perot cavity
described by a scattering matrix $S$ which can be deduced from the
two elementary matrices. The Casimir force then results from the
difference of radiation pressures exerted onto the inner and outer
sides of the mirrors by the vacuum field
fluctuations~\cite{JaekelJP91}. Equivalently, the Casimir free
energy can be derived from the frequency shifts of all vacuum field
modes due to the presence of the cavity.

The same discussion can be extended to the geometry of two plane and
parallel mirrors aligned along the axis $x$ and $y$, described by
specular reflection and transmission amplitudes which depend on
frequency $\omega$, the transverse vector $\bk \equiv \left(
k_x,k_y\right)$ and the polarization $p=\TE,\TM$. A few points have
to be treated with care when extending the derivation from
1-dimensional space to 3-dimensional space~: evanescent waves
contribute besides ordinary modes freely propagating outside and
inside the cavity; dissipation has to be accounted
for~\cite{GenetPRA03}. The properties of the evanescent waves are
described through an analytical continuation of those of ordinary
ones, using the well defined analytic behavior of the scattering
amplitudes. At the end of this derivation, this analytic properties
are also used to perform a Wick rotation from real to imaginary
frequencies.

The sum of all phaseshifts leads to the expression of the Casimir
free energy $\calF$
\begin{eqnarray}
\label{CasimirFreeEnergy} &&\calF = \sum_\bk \sum_p k_\B T
\sum_m{}^\prime \ln d(i\xi_m,\bk,p) \\ &&d = 1 - r e^{ -2\K L }
\,,\quad \xi_m \equiv \frac{2\pi m k_\B T}\hbar \,,\quad \K\equiv
\sqrt{\bk^2+\frac{\xi^2}{c^2}} \nonumber
\end{eqnarray}
$\sum_\bk \equiv A \int\frac{\dd^2\bk}{4\pi^2}$ is the sum over
transverse wavevectors with $A$ the area of the plates, $\sum_p$ the
sum over polarizations and $\sum_m^\prime$ the Matsubara sum (sum
over positive integers $m$ with $m=0$ counted with a weight
$\frac12$); $d$ is the denominator describing cavity resonances;
$r\equiv r_1 r_2$ is the product of the reflection amplitudes of the
mirrors as seen by the intracavity field; $\xi$ and $\K$ are the
counterparts of frequency $\omega$ and longitudinal wavevector $k_z$
after the Wick rotation.

This expression reproduces the Casimir ideal formula in the limits
of perfect reflection $r \rightarrow 1$ and null temperature $T
\rightarrow 0$. But it is valid and regular at thermal equilibrium
at any temperature and for any optical model of mirrors obeying
causality and high frequency transparency properties. It can thus be
used for calculating the Casimir force between arbitrary mirrors, as
soon as the reflection amplitudes are specified. These amplitudes
are commonly deduced from models of mirrors, the simplest of which
is the well known Lifshitz model
\cite{LifshitzJETP56,DzyaloshinskiiUspekhi61} which corresponds to
semi-infinite bulk mirrors characterized by a local dielectric
response function $\varepsilon (\omega)$ and reflection amplitudes
deduced from the Fresnel law. In principle, the expression
(\ref{CasimirFreeEnergy}) can still be written in terms of
reflection amplitudes even when the optical response of the mirrors
can no longer be described by a local dielectric response function.

\section{The non specular scattering approach}

The scattering formalism can be generalized one step further to
calculate the Casimir force between stationary objects with
arbitrary geometries. Now the scattering matrix $\calS$ is a larger
matrix accounting for non-specular reflection and mixing different
wavevectors and polarizations while preserving frequency. Of course,
the non-specular scattering formula is the generic one while the
specular limit can only be an idealization.

The Casimir free energy can be written as a generalization of
equation (\ref{CasimirFreeEnergy})
\begin{eqnarray}
\label{CasimirFreeEnergyNS} &&\calF = k_\B T \sum_m{}^\prime \,\Tr
\ln \calD (i\xi_m) \\ &&\calD = 1 - \calR_1 \exp^{ -\calK L }
\calR_2 \exp^{ -\calK L } \nonumber
\end{eqnarray}
The symbol $\Tr$ refers to a trace over the modes at a given
frequency. The matrix $\calD$ is the denominator containing all the
resonance properties of the cavity formed by the two objects 1 and 2
here written for imaginary frequencies. It is expressed in terms of
the matrices $\calR_1$ and $\calR_2$ which represent reflection on
the two objects 1 and 2 and of propagation factors $\exp^{-\calK
L}$. Note that the matrices $\calD$, $\calR_1$ and $\calR_2$, which
were diagonal on the basis of plane waves when they described
specular scattering, are no longer diagonal in the general case of
non specular scattering. The propagation factors remain diagonal in
this basis with their diagonal values written as in
(\ref{CasimirFreeEnergy}). Clearly the expression
(\ref{CasimirFreeEnergyNS}) does not depend on the choice of a
specific basis. But it may be written in specific basis fitting the
geometry under study.

The multiple scattering formalism has been used in the past years by
different groups using different notations (see as examples
\cite{EmigJSM08,KennethPRB08,MiltonJPA08}). Numerous applications
have been considered and some of them are discussed in the next
section.

\section{Applications to different geometries}

Various geometries can be studied beyond the PFA by using the
general expression (\ref{CasimirFreeEnergyNS}). For example, one can
study plane or spherical plates, flat, rough or nanostructured
surfaces, atoms, molecules or nanoparticles, as well as different
combinations of these possibilities.

The first applications were devoted to the effect of surface
roughness on the normal Casimir
force~\cite{GenetEPL03,MaiaNetoEPL05,MaiaNetoPRA05}. The case of
corrugated plates is also interesting, in particular because it
gives rise to lateral forces when the corrugations are shifted with
respect to each other~\cite{RodriguesPRL06,RodriguesPRA07,ChiuPRB09}
and to torques when they are misaligned~\cite{RodriguesEPL06}. In
the geometry of a sphere above a grating, the normal Casimir force
is affected in a manner which can be understood quantitatively
by using the non specular scattering approach%
~\cite{ChanPRL08,LambrechtPRL08,LambrechtNat08,BaoPRL10}.

Applications have also been developed for the study of atoms in the
vicinity of corrugated plates~\cite{MessinaPRA09,ContrerasPRA10}.
Examples of such applications involve the use of Bose-Einstein
condensates (BEC) as probes of vacuum affected by the proximity of a
grooved surface~\cite{DalvitPRL08}, localization of matter waves in
the disordered vacuum above a rough plate~\cite{MorenoPRL10}, and
the driving of quantized vortices in a BEC by a contactless tranfer
of angular momentum through the mediation of vacuum
fluctuations~\cite{ImpensEPL10}. All these applications involve
Casimir forces or torques beyond the regime of validity of the PFA.

Efforts have also been devoted to the use of multiple scattering
method to obtain explicit evaluations of the Casimir force in the
plane-sphere geometry.  Such calculations have first been performed
for perfectly reflecting mirrors~\cite{MaiaPRA08}. They have then
been done for the more realistic case of metallic mirrors described
by a plasma model dielectric function~\cite{CanaguierPRL09}. More
recently, it has become possible to make calculations which treat
simultaneously plane-sphere geometry and non zero temperature, with
dissipation taken into account~\cite{CanaguierPRL10}. In these
calculations, the reflection matrices are written in terms of
Fresnel amplitudes for plane waves on the plane mirror and of Mie
amplitudes for spherical waves on the spherical mirror. The
scattering formula is then obtained by writing transformation
formulas from the plane waves basis to the spherical waves basis and
conversely. The energy takes the form of an exact multipolar formula
labeled by a multipolar index $\ell$. When doing the numerics, the
expansion is truncated at some maximum value $\ell_\max$, which
degrades the accuracy of the resulting estimation for very large
spheres $x\equiv L/R<x_\min$ with $x_\min$ proportional to
$\ell_\max^{-1}$.

The results of these calculations may be compared to the only
experiment devoted to the study of PFA in the plane-sphere
geometry~\cite{KrausePRL07}. In this experiment, the force gradient
$G$ was measured for various radii of the sphere and the results
were used to constrain the value of the slope at origin $\beta_G$ of
the function
\begin{eqnarray}
\rho_G(x)=\frac{G}{G^\PFA}=1+\beta_G x+O(x^2) \quad,\quad x\equiv
\frac LR
\end{eqnarray}
where $x$ is the already discussed aspect ratio which characterizes
the plane-sphere geometry. The constraint obtained in this
experiment is read $\vert\beta_G\vert<0.4$. Its comparison with the
theoretical value obtained for the slope $\beta_G$ by interpolating
at low values of $x$ the numerically evaluated $\rho_G$ reveals a
striking difference between the cases of perfect and plasma mirrors.
The slope $\beta_G^\perf\sim-0.48$ obtained for perfect mirrors is
indeed not compatible with the experimental bound. In contrast, the
slope $\beta_G^\Gold\sim-0.21$ obtained for more precisely described
gold mirrors is significantly smaller and, as a result, compatible
with the experimental bound.

The effect of temperature is also correlated with that of
plane-sphere geometry. The first calculations accounting
simultaneously for plane-sphere geometry, temperature and
dissipation~\cite{CanaguierPRL10} show several striking features.
The factor 2 between two planes at long distances calculated with
Drude and plasma models is reduced to a factor below 3/2 in the
plane-sphere geometry. Then, PFA underestimates the Casimir force
within the Drude model at short distances, but overestimates it at
all distances for the perfect reflector and plasma model. If the
latter feature were conserved for the aspect ratios met in the
experiments, the actual values of the Casimir force calculated
within plasma and Drude model could be closer than what PFA
suggests. This would affect the comparison of Casimir measurements
with theory, which is still based on calculations using PFA.

We finally refer to the study of the Casimir interaction between a
dielectric nanosphere and a metallic plane~\cite{CanaguierPRA11}.
The known Casimir-Polder formula is recovered at the limit of small
nanospheres, which may be thought of as large atoms. Meanwhile an
expression that takes into account the finite size of the sphere is
found, which behaves better at small distances than the
Casimir-Polder formula. This opens the way to new studies devoted to
the optomechanics of nanoobjects in the vacuum modified by the
proximity of a surface.

\section*{Acknowledgments}
The authors thank A. Canaguier-Durand, A. G\'erardin, R. Gu\'erout,
J. Lussange, R.O. Behunin, I. Cavero-Pelaez, D.A.R. Dalvit, C.
Genet, G.L. Ingold, F. Intravaia, M.-T. Jaekel, P.A. Maia Neto, V.V.
Nesvizhevsky, A.Yu. Voronin, for contributions to the work reviewed
in this paper, and the ESF Research Networking Programme CASIMIR
(www.casimirnetwork. com) for providing excellent opportunities for
discussions on the Casimir effect and related topics.

\end{document}